\documentclass[numberedappendix]{emulateapj}
\slugcomment{Accepted by The Astrophysical Journal Supplement Series}
\shorttitle{Gini: a stable measure of galaxy structure?}
\shortauthors{T. Lisker}
\begin{document}
 
\title{Is the Gini coefficient a stable measure of galaxy structure?}

\author{Thorsten Lisker}
\affil{Astronomisches Rechen-Institut, Zentrum f\"ur Astronomie der
  Universit\"at Heidelberg (ZAH), M\"onchhofstra\ss e 12-14, D-69120
  Heidelberg, Germany}
\email{TL@x-astro.net}

\begin{abstract}
The Gini coefficient, a non-parametric measure of  galaxy morphology, has
recently taken up an important role in the automated identification of galaxy
mergers. I present a critical assessment of its stability, based on a
comparison of HST/ACS imaging data from the GOODS and UDF surveys. Below a
certain signal-to-noise level, the Gini coefficient depends strongly
on the signal-to-noise ratio, and thus becomes useless for distinguishing
different galaxy morphologies. Moreover, at all signal-to-noise levels the
Gini coefficient shows a strong dependence on the choice of aperture within
which it is measured.
Consequently,  
quantitative selection criteria involving the Gini coefficient, such as a
selection of merger candidates, cannot always be straightforwardly applied to
different datasets. 
 I discuss whether these effects could have affected previous studies that
 were based on the Gini coefficient, and establish signal-to-noise limits
 above which measured Gini values can be considered reliable.
\end{abstract}
 
 

\section{Introduction and Motivation}
 \label{sec:intro}

The ever increasing amount of galaxies probed by imaging surveys such as
the Sloan Digital Sky Survey \citep[SDSS,][]{sdssedr}, the Great
Observatories Origins Deep Survey \citep[GOODS,][]{goods}, or the Hubble
Space Telescope (HST) Ultra Deep Field \citep[UDF,][]{udf}
triggered the need for automated galaxy classification methods. Ideally,
these should not only be faster in classifying thousands of galaxies
than any visual examination, but should be more robust and less subjective
than the latter. 

An important aspect is the evolution of galaxy structure over cosmic time:
deep HST surveys such as GOODS or UDF reveal a huge amount of peculiar or
irregular galaxies \citep[see, e.g.,][]{cow95a,elm04b,con08} that do not fit into
the standard Hubble scheme, nor 
into the class of irregular galaxies as defined for the local
Universe. This is owed to the fact that more distant galaxies are seen in
a significantly younger stage, in which starburst events and mergers occur
at a larger rate \citep[e.g.][]{cow95b,mad98,bell06a}. Moreover, at higher
redshifts, bluer rest-frame 
wavelengths are probed, giving emphasis to the distribution of the
young stellar component.
Although trust in visual classification remains for both low-redshift
\citep[e.g.][]{sch07b} and high-redshift datasets \citep[e.g.][]{ferlis05},
the hope was that
new structural classification schemes  would also be suitable to
describe the various kinds of peculiar galaxies in an appropriate way
\citep{abr94,abr96b}, and to 
separate them from the more familiar galaxy classes.

``Non-parametric'' structural measures that have been developed in this context
 are the concentration index \citep[$C$,][]{abr94,ber00,tru01,gra01,con03a},
 the asymmetry index 
 \citep[$A$,][]{abr96a,con97,kor98,con00}, and the clumpiness index
 \citep[$S$,][]{con03a}. These three are known as the CAS parameters
 \citep{con03a}. 
  Two further parameters that are frequently used are the Gini coefficient
 \citep{gini,abr03,lot04} and the $M_{20}$ parameter \citep{lot04}. Gini
describes the distribution of flux values among the pixels of an object's
 image, while $M_{20}$ quantifies the distribution of the brightest 20\% of
 pixels. As
 opposed to the CAS parameters, neither Gini nor $M_{20}$ require the center
 of a galaxy to be defined, and thus do not need to assume that a
 well-defined visible center exists at all. All these parameters are commonly
 dubbed ``non-parametric'', since they do not rely on certain model parameter
 fits, such as the S\'ersic index \citep{ser63} in models of radial surface
 brightness profiles. A more appropriate term might therefore be ``model-independent''.

One particular application of these model-independent measures
 was the
identification of galaxy mergers.
Studying their abundance and
properties is crucial to understand the origin of the most massive
galaxies in the Universe today \citep[e.g.][]{bell06b}, of which many
must have formed within the last gigayears \citep{bell04, ferlis05}.
\citet{con03b} selected major galaxy mergers as objects with
a rest-frame $B$-band asymmetry index of $A\geq 0.35$, and used it to derive
merger fractions and subsequently galaxy merger rates out to a redshift of
$z\approx 3$.
By applying this structural selection to submillimeter-detected
galaxies, \citet{con03c} concluded that many of these are undergoing a major
merger. 
 \citet{lot06} studied the fraction of major mergers and minor merger
 candidates among high-redshift star-forming galaxies, using a classification
 based on Gini, $M_{20}$, and concentration. A major merger selection
 criterion within the  two-dimensional parameter space of Gini ($G$) and
$M_{20}$ was established by \citet{lot08} as
$G > -0.14\cdot M_{20} + 0.33$, enabling them to study the evolution of the
 galaxy merger rate.

The Gini coefficient
was found to correlate strongly with stellar mass \citep{zam07}. In fact,
\citeauthor{zam07} claim that Gini traces
better the overall structure of a galaxy than any other
morphological parameter of their study.
Apart from describing the overall structure, the Gini coefficient was also
used to identify substructure in galaxies: \citet{celeste} presented a
preliminary method to automatically identify large bars, using the radial
variation of Gini within a galaxy.

The Gini coefficient therefore plays an important role in present-day attempts
to understand the formation and evolution of galaxy structure, and in
particular, to unveil the formation process of massive galaxies.
Despite simulations and a direct comparison of GOODS and UDF images
\citep{lot04,lot06}, which were done in order to
quantify how Gini and other parameters depend on the signal-to-noise ratio and
angular size resolution,
the stability of the Gini coefficient has not been fully examined yet. In
what follows, I
present a critical assessment of the stability of Gini with respect
to the signal-to-noise ratio and the choice of aperture.

%
%
%


\section{Data and measurements}
 \label{sec:data}

The HST UDF \citep{udf} and GOODS \citep{goods} ACS imaging data are ideally
suited to investigate the dependence of the Gini coefficient on
the signal-to-noise ratio ($S/N$), and hence the depth, of an image.
 These two
datasets were observed with the same instrumental setup, used very similar
data reduction procedures, and differ only in their total exposure
time. Moreover, along with the UDF images, astrometrically
registered images of the corresponding GOODS area are
provided\footnote{http://www.stsci.edu/hst/udf/goods}, which can be
ideally used for a direct investigation of the effect of image depth on
the analysis of galaxy structure.

The measurements described below were performed on the $i_{775}$ images.
As initial galaxy sample, I use the $i_{775}$ UDF source catalog
 \citep{udf}. Sources that had been identified manually \citep[cf.][]{udf}
 were excluded, since these are not included in the available segmentation
 maps\footnote{ftp://udf.eso.org/archive/pub/udf/acs-wfc}, which denote
 the spatial extent of each object. Sources with stellarity$\ge0.8$ were
 excluded, following \citet{coe06}, leaving 8276 objects.

Both GOODS and UDF suffer from residual (i.e.\ non-zero) background flux,
which typically reaches a level of 1/5 of the noise RMS in GOODS and 1/20 in
UDF, in the $i_{775}$ band. This background, along with a value of the
noise RMS, was determined
individually for each galaxy as a single value that corresponds to the
median pixel value within a $21''\times 21''$-box, applying five iterations of
clipping outliers at 2.3 standard deviations. All 
galaxies were masked in this process.

 For each galaxy, I determined a ``Petrosian semimajor axis'' (hereafter
 Petrosian SMA, $a_{\rm Petro}$), i.e.,
 in the calculation of the Petrosian radius \citep{pet76},  I use
 ellipses instead of circles \citep[cf.][]{lot04,p3}. The elliptical shape
 and source center
 of each object were adopted from the UDF source catalog, and kept fixed
 during the process. The Petrosian SMA was defined as the semimajor axis $a$
 at which the local intensity falls below one fifth of the
 average intensity within $a$. The local intensity is
 measured as the average intensity within an elliptical annulus
 reaching from $0.9 a$ to $1.1 a$, applying five iterations of
 clipping outlying pixel values at 2.3 standard deviations, and masking
 neighbouring objects.
 For 181 objects, the Petrosian SMA determination did not converge or
 yielded a too large value based on visual inspection. These were
 excluded, as well as 92 further objects that comprised less than 28
 pixels each (corresponding to a circle with a radius of 3 pixels). The
 final working sample thus contains 8003 galaxies.
 The total flux of each galaxy was measured within $a = 2\,a_{\rm Petro}$.
For each galaxy, the Petrosian radius was calculated on the UDF image only,
and was then used on both the UDF and the GOODS image.

The Gini coefficient was calculated 
following the definition of \citet{lot04,lot08},
\begin{equation}
G = \frac{ \sum^{N_{pix}}_i (2i -
N_{pix} -1) |f_i|}{(N_{pix}-1) \sum^{N_{pix}}_i |f_i|}
\end{equation}
where $N_{pix}$ is the number of pixels in the image, $f_i$ is the
pixel flux, and $|f_1| \leq |f_2| \leq |f_3| \leq ... \leq |f_{N_{pix}}|$.  As
discussed by \citet{lot04}, the absolute values of $f_i$ are required in
order to preserve the correct structure at low flux levels, where
noise can result in negative values of $f_i$ after background
subtraction.
For this calculation, all pixels within a given aperture were used.
Gini was measured for the UDF and GOODS images on three different
elliptical apertures: $a=2/3\,a_{\rm Petro}$ (``2/3-Petrosian aperture''), $a=1\,a_{\rm Petro}$
(``Petrosian aperture''), and $a=3/2\,a_{\rm Petro}$ (``1.5-Petrosian aperture'').


\section{The Gini coefficient's dependencies}
 \label{sec:gini}

\begin{figure}
  \epsscale{1.1}
  \plotone{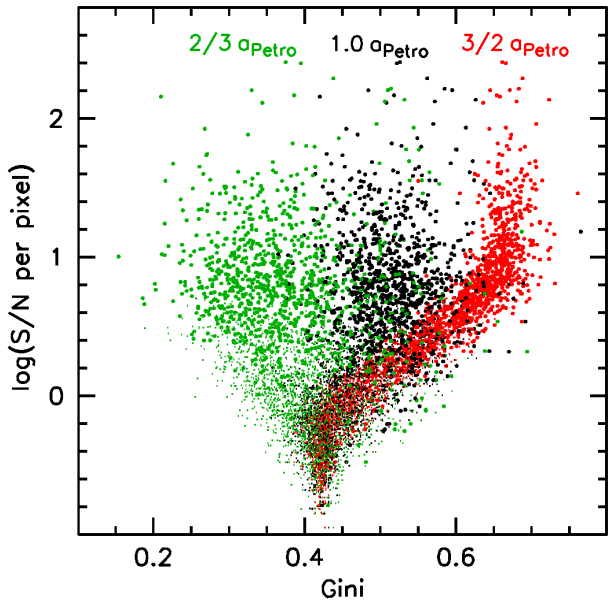}
  \caption{{\bf Gini for different apertures.}
    Logarithmic signal-to-noise ratio per pixel in UDF, measured within the Petrosian
    aperture, versus the Gini coefficient of 
    the galaxies in UDF. Gini is shown for the three different apertures:
    $a=2/3\,a_{\rm Petro}$ (green or grey symbols), $a=1\,a_{\rm Petro}$
    (black), and $a=3/2\,a_{\rm Petro}$ (red or grey). Large symbols 
    represent the galaxies in the good quality subsample, i.e., having
    $a_{\rm Petro}\geq 10$ pixels and $m_i\le 
    26.7$ mag. Small symbols are used for the other galaxies.
    For a given galaxy, the $S/N$ per pixel
    is defined as the average flux per pixel, divided by the noise RMS
    value calculated in a box centered on the
    galaxy's position.
  }
  \label{fig:aperture}
\end{figure}

\begin{figure}
  \epsscale{1.0}
  \plotone{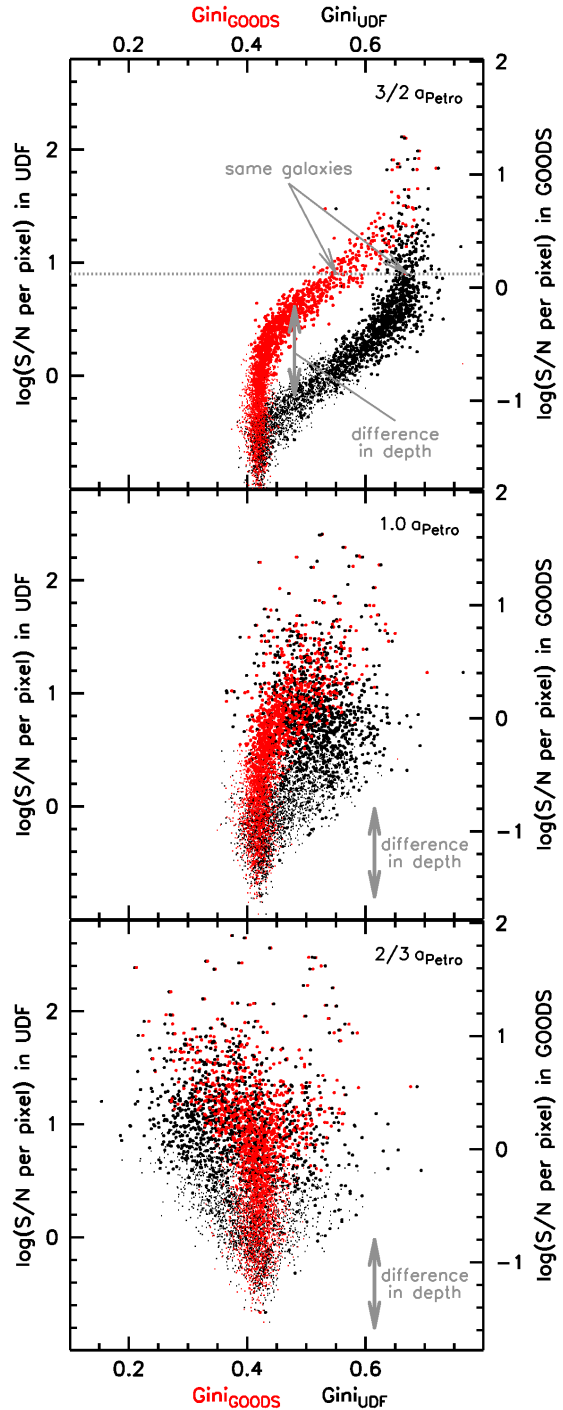}
  \caption{{\bf Gini for different image depth.}
    Logarithmic signal-to-noise ratio per pixel versus the Gini coefficient of
    the same galaxies in UDF (black symbols) and GOODS (red or
    white). \emph{Top:} 1.5-Petrosian aperture, \emph{middle:}  
    Petrosian aperture,  {\emph bottom:} 2/3-Petrosian
    aperture.
    The median difference between the noise RMS of UDF
    and GOODS is indicated as arrow in each panel.
For a given galaxy, the Gini values of UDF and GOODS lie on a
horizontal line in each panel, since the galaxy's $S/N$ value in UDF
(left ordinate) was used for plotting, whereas the right ordinate
has simply been scaled according to the depth difference to represent the
    GOODS $S/N$ values.
  }
  \label{fig:depth}
\end{figure}

The strong dependence of the Gini coefficient on the aperture within which it
is measured is shown in Fig.~\ref{fig:aperture}. For each of the three
apertures, Gini occupies a completely different range of values for 
galaxies with a reasonably high $S/N$.
The importance of this aperture dependence becomes clear when considering the
fact that different authors keep using different apertures in their
computation of model-independent morphology measurements: while, for example,
\citet{con03a} and \citet{con00,con08}
use 1.5 circular Petrosian radii,
\citet{lot04,lot06,lot08} use 1.0 Petrosian SMA.

Moreover, the figure implies a $S/N$-dependence of the Gini coefficient: from
the high-$S/N$ values, a transition is observed towards a common Gini value
of 0.42 for all apertures at very low $S/N$, where noise dominates over the
actual signal.
However, from Fig.~\ref{fig:aperture} alone it is not possible to judge
whether this transition is a direct dependence on $S/N$, or whether it simply
means that the Gini coefficient takes a different value for galaxies of
different surface brightness, which could be useful to distinguish between
different galaxy classes.

Consequently, a direct examination of the effects of image depth is
necessary. It is made possible through a comparison of the very same galaxies
and apertures in the UDF and GOODS images, which differ only by the total
exposure time. 
This comparison is shown
in Fig.~\ref{fig:depth} for the three different apertures. The arrow in
each panel indicates the difference in depth between UDF
and GOODS (1.95 mag), calculated from the median of the individual noise RMS
measurements.\footnote{Effects of the drizzling process on the noise
  measurement \citep[cf.][]{lot06} are not taken into account, since they
  would only shift the diagram axes but not alter the comparison of GOODS and
  UDF images.}
For a given galaxy, the Gini values of UDF and GOODS lie on a
horizontal line in each panel, since the galaxy's $S/N$ value in UDF
(left ordinate) was used for plotting, whereas the right ordinate
has simply been scaled according to the depth difference to represent the
     GOODS $S/N$ values.

Obviously, the depth difference equals the offset of the Gini
distributions for the two images. For the example of the 1.5-Petrosian aperture
(top panel in Fig.~\ref{fig:depth}), this
means that the transition of Gini values from $\sim0.65$ to  $\sim0.4$ is
determined solely by the $S/N$, and not by different surface brightnesses of
galaxies --- the latter do not change between UDF and GOODS.
Similar structures, though not as clearly defined, can be seen for the
Petrosian and the 2/3-Petrosian aperture (middle and bottom panel). Gini thus
shows a strong dependence on the $S/N$ of an image.

\begin{figure}
  \epsscale{1.0}
  \plotone{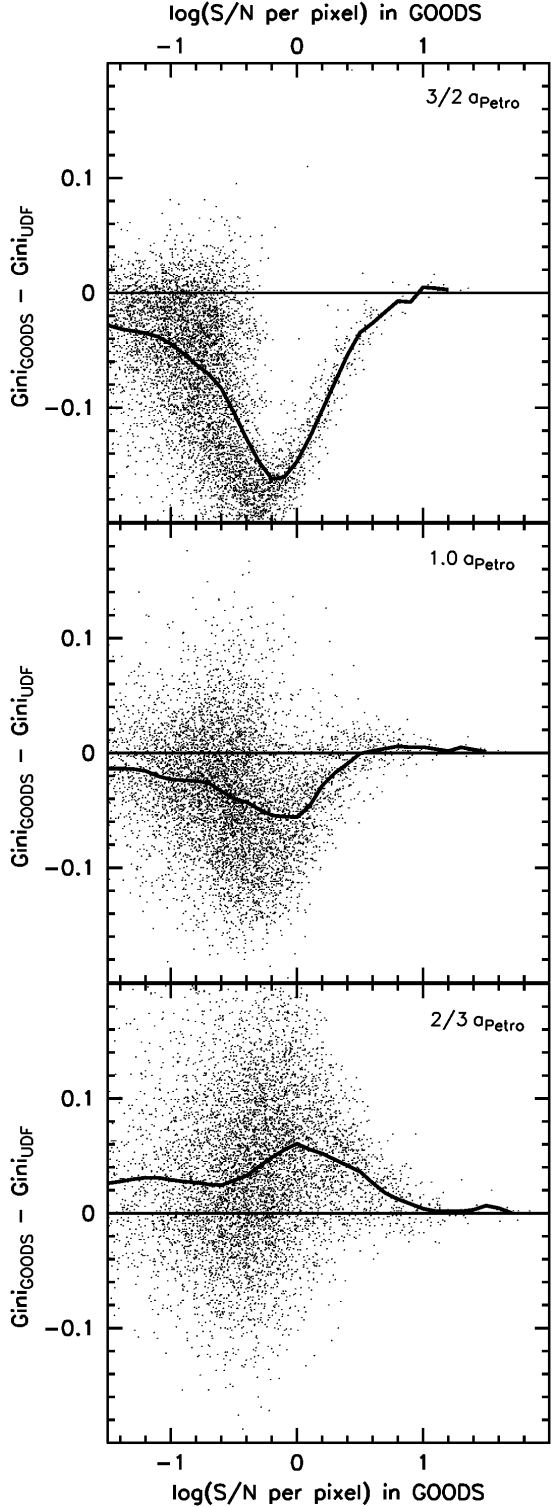}
  \caption{{\bf Gini versus $S/N$.}
    Logarithmic signal-to-noise ratio per pixel in GOODS versus the difference
    of the Gini coefficient in GOODS and UDF. The black line gives the average
    Gini difference.
    {\emph Top:} 1.5-Petrosian aperture, \emph{middle:} Petrosian aperture,
    \emph{bottom:} 2/3-Petrosian aperture.
  }
  \label{fig:difference}
\end{figure}

\begin{figure}
  \epsscale{1.1}
  \plotone{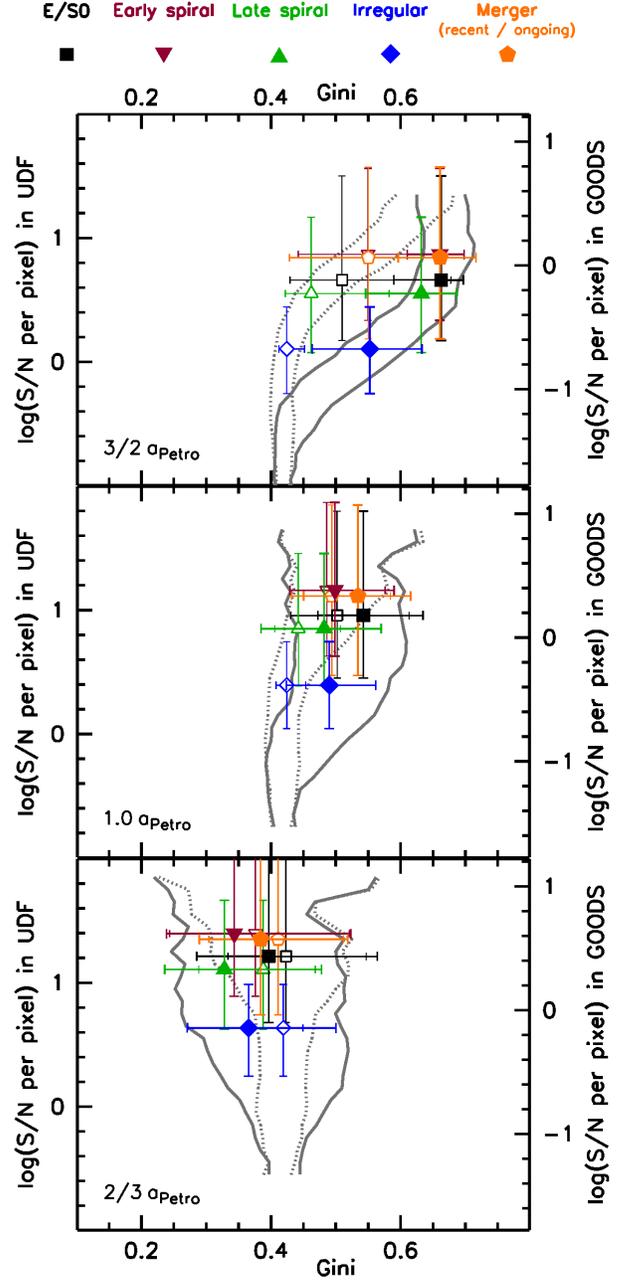}
  \caption{{\bf Gini for different galaxy classes.}
    Logarithmic signal-to-noise ratio per pixel versus the Gini coefficient
    for different galaxy classes (see text), as indicated above the
    diagram. Shown are 
    the median values in UDF (filled symbols) and GOODS (open symbols); error
    bars represent the 5th and 95th percentile.
    Contours indicate the 5\%-to-95\% range of Gini values (cf.\
    Fig.~\ref{fig:depth}) in UDF (grey solid lines) and GOODS (grey dotted
    lines).
For a given galaxy class, the Gini values of UDF and GOODS lie on a
horizontal line in each panel, since the UDF values
(left ordinate) were used for plotting, whereas the right ordinate
has simply been scaled according to the depth difference to represent the
    GOODS $S/N$ values.
    {\emph Top:} 1.5-Petrosian aperture, \emph{middle:} 
    Petrosian aperture, \emph{bottom:} 2/3-Petrosian aperture.
  }
  \label{fig:class}
\end{figure}

In order to have a more quantitative measure of the systematic $S/N$
effect on the Gini coefficient, Fig.~\ref{fig:difference} shows the
difference between the GOODS and the UDF Gini value for each galaxy,
versus the $S/N$. The same comparison was presented by \citet{lot06} for
their sample and aperture. From this, they decided to restrict their
analysis to galaxies with $S/N$ per pixel of $S/N_{p.p.} \gtrsim 2.5$ in
GOODS, corresponding to $log(S/N_{p.p.})\gtrsim 0.4$. This is in agreement
with the middle panel of Fig.~\ref{fig:difference}, which illustrates that
the difference between the GOODS and the UDF values for the Petrosian
aperture is very small at this $S/N$ level.

 The Gini distributions of different galaxy classes in GOODS
and UDF are shown in Fig.~\ref{fig:class}, again for the three elliptical
apertures. Galaxy classification was done visually for all 1019 objects
with
$a_{\rm Petro}\geq 10$ pixels and $m_i\le 26.7$ mag, hereafter referred to
as the good quality subsample.\footnote{At this magnitude, approximately
  90\% of the galaxies have $S/N_{p.p.}\geq 1$ in UDF
  within
  the Petrosian aperture.
  For a given galaxy, the $S/N$ per pixel is defined as the average flux
  per pixel, divided by the noise RMS value calculated in a box centered on the
    galaxy's position.}
Since many, if not most
galaxies in deep surveys such 
as UDF are not well represented by any 
class of the Hubble scheme, a number of additional classes can be defined
that are a more ``honest'' representation of the diversity of galaxies at
higher redshifts, such as chain galaxies, clump clusters
\citep[see][]{elm04a,elm04b}, 
and compact or high-surface brightness irregular galaxies. While this
detailed classification will be used for future projects, the intention
here is to get an idea of how the standard galaxy classes distribute (see
the discussion),
along with the combined class of candidates for ongoing and recent
mergers. Therefore, only these are shown in the figure, with their median
value and the corresponding interval that reaches from the 5th to the 95th
percentile.

It can be seen that E/S0 galaxies and possible mergers, which typically
have comparably high surface brightnesses, have similar Gini
values. Recall that Gini is not sensitive to the actual location of galaxy
pixels; thus, any asymmetry or irregularity in the spatial distribution
does not enter the Gini calculation. Early-type spirals have a lower value
at least for the 2/3-Petrosian and the Petrosian aperture, and late-type
spirals have again somewhat lower values. The irregulars have a
significantly lower $S/N$ on average, causing their comparison with the other
classes to be affected strongly by $S/N$ effects: their Gini
values are larger than those of the spirals for the 2/3-Petrosian
aperture (bottom panel of Fig.~\ref{fig:class}), about equal
for the Petrosian aperture (middle), and
significantly lower for the 1.5-Petrosian aperture (top). As an aside, note the
large overlap of all galaxy types for all 
apertures --- while it might be possible to perform
``non-parametric'' quantitative galaxy classification in a statistical way, it is barely
possible to obtain \emph{clean} samples of each galaxy class.


\section{Discussion}
\label{sec:discuss}

\subsection{Merger selection}
Determining the frequency of major galaxy mergers at different epochs is a
crucial step towards understanding the origin of present-day massive
galaxies. In this context, \citet{lot08} identified mergers within
the parameter space of the Gini coefficient $G$ and $M_{20}$ \citep[which
  measures the distribution of the brightest 20\% of a galaxy's pixels,
  see][]{lot04} by selecting galaxies with $G > -0.14\cdot M_{20} +
0.33$. However,  
\citet{con08} noted that, when applying this criterion\footnote{
  \citet{con08} actually used $G > -0.115\cdot M_{20} + 0.384$ from the
  submitted version of \citet[arXiv:astro-ph/0602088v1]{lot08}.
}
to their galaxy
sample, the majority of selected objects actually appear to be normal
galaxies according to their CAS (concentration, asymmetry, clumpiness)
classification criteria. They concluded that the Gini/$M_{20}$ criterion
might not be selecting true mergers, or might be more sensitive to
specific merger phases.

From the
analysis presented here, it is immediately clear that this
apparent discrepancy is caused by the strong dependence of Gini on the
chosen aperture: while \citet{lot08} used 1.0
Petrosian SMA, \citet{con08} chose 1.5 Petrosian radii for their
aperture. Since the average Gini value increases significantly with
increasing aperture (Fig.~\ref{fig:aperture}), \citeauthor{con08}'s
application of the Gini/$M_{20}$ criterion  selected many
more objects as ``mergers'' than it would have
with \citeauthor{lot08}'s aperture.\footnote{
  Hypothetically, the variation of $M_{20}$ with aperture
  size could happen to be such that it counterbalances that of
  Gini. However, this is not the case: the variation of $M_{20}$ with
  aperture size is found to be significantly smaller than what would be
  necessary for the merger criterion to remain valid on average \citep[also
    see][]{lot06}.
}
Furthermore, while
\citet{lot08} used elliptical apertures in order to avoid the
inclusion of many noise-dominated pixels \citep{lot04},
\citet{con08} used circular apertures, thereby again increasing the average
Gini values of their galaxies.

\citeauthor{con08} quote an average Gini
value of 0.71 for their UDF galaxy sample, measured in the $z_{850}$
band. In the present study, the average value for the circular aperture of the same size, but
measured in $i_{775}$, is 0.64, using only the good quality subsample. The
corresponding values for the 
elliptical apertures are 0.38 for the 2/3-Petrosian aperture, 0.52 for the
Petrosian aperture, and 0.62 for the 1.5-Petrosian aperture.



\subsection{Choice of  aperture}

The signal-to-noise ratio of a galaxy is intrinsically tied to its surface
brightness. 
Nevertheless, as emphasized in Sect.~\ref{sec:gini}, the systematic change
of the Gini coefficient from high to low $S/N$ values when considering the
1.5-Petrosian aperture is a true $S/N$-effect (see Fig.~\ref{fig:depth}) and is not
caused by varying surface brightness. This, in turn, means that each
$S/N$ level defines its own specific range of possible Gini values that
galaxies at this $S/N$ level can take. Therefore, a galaxy could only be properly compared to
other galaxies if those are at about the same $S/N$ level --- a rather
alarming result.
However, as can be seen from  Fig.~\ref{fig:depth}, this effect is far
less strong for the 2/3-Petrosian and the Petrosian aperture. For these
apertures, the range of Gini values covered by different galaxies is much
larger with respect to the systematic $S/N$-effect --- although the latter
is still clearly recognizable through the difference
in depth between UDF and GOODS.

The maximum systematic
deviation of the Gini value that one finds acceptable can of course
be subjective. Since the Gini coefficient is intended to be used to measure
galaxy morphology, a useful criterion would seem that the separation between
different galaxy classes be larger than the systematic
$S/N$ effect.
An acceptable limit for the systematic $S/N$ effect
can, for example, be defined as half the difference between the median Gini
values of E/S0s and late-type spirals in UDF (see
Fig.~\ref{fig:class}). This would be 
0.035 for the 2/3-Petrosian aperture,
0.03 for the Petrosian aperture, and
0.015 for the 1.5-Petrosian aperture.
The corresponding $S/N$-restrictions from
Fig.~\ref{fig:difference} would then be
$log(S/N_{p.p.})\gtrsim 0.5$ for the 2/3-Petrosian aperture,
$log(S/N_{p.p.})\gtrsim 0.2$ for the Petrosian aperture, and
$log(S/N_{p.p.})\gtrsim 0.7$ for the 1.5-Petrosian aperture.
When taking into account the effects of drizzling on the noise in GOODS
\citep{cas00,lot06}, these limits would decrease by 0.2.
Note that the $S/N$ values are calculated as the average $S/N$ of all
pixels within a given aperture --- it is therefore easier to obtain a larger $S/N$ value
for smaller apertures, which contain fewer faint pixels.

The clear
conclusion  is that for the vast majority of
galaxies, the Gini coefficient is strongly affected by the $S/N$ when
using the 1.5-Petrosian aperture: for the GOODS images, only 2\% of the galaxies in the good
quality subsample meet the above criterion. For the 2/3-Petrosian and the
Petrosian aperture, this fraction is significantly larger, namely 24\% and
30\%, respectively. It is therefore recommended to use the Petrosian
aperture for calculating the Gini coefficient, confirming the approach of
\citet{lot04,lot06}.

\subsection{Assessment of previous studies}

With the criteria established above, it is possible to assess whether past
studies that used the Gini coefficient for galaxy classification might
have been significantly affected by its systematic behaviour with
$S/N$. As mentioned already in Sect.~\ref{sec:gini}, in their studies of
intermediate and high redshift galaxy morphology and merger fractions,
\citet{lot06,lot08} selected objects having $log(S/N_{p.p.})\gtrsim
0.4$ within the Petrosian aperture\footnote{Furthermore, \citet{lot06}
  corrected the $S/N$ values in GOODS for the effects of drizzling,
  thereby lowering the $S/N$ values and making their selection even more
  restrictive.}, for which there is almost
no difference between the Gini values in GOODS and UDF
(Fig.~\ref{fig:difference}, middle panel). The same selection was
adopted by \citet{pie07} in their study of AGN host galaxy morphologies.

The study of \citet{con08} of galaxy structures and merger fractions
in UDF used galaxies with $z_{850}<27.0$ mag, and adopted a circular
aperture of 1.5 Petrosian radii, as mentioned previously. When requiring
that the above criterion of
$log(S/N_{p.p.})\gtrsim 0.7$ for the 1.5-Petrosian aperture in
$i_{775}$ be met by more than 50\% of the galaxies with magnitudes similar
to the limiting
magnitude, the latter would have to be approximately $i_{775}\le 24.5$ mag
(and leave only very few galaxies in a so selected sample). Depending on
galaxy type and color, the limiting magnitude in $z_{850}$ would differ by
several tenths of a magnitude, but still be more than two magnitudes
brighter than the selection of \citet{con08}. Consequently, the Gini
values that \citeauthor{con08} obtained for the majority of their sample
must be regarded as being strongly affected 
by $S/N$ effects. If they had 
chosen the Petrosian aperture instead, the above criterion of
$log(S/N_{p.p.})\gtrsim 0.2$ would have been met by about 80\% of
galaxies with magnitudes around $i_{775}\le 27.0$.
Furthermore, it needs to be pointed out
again that their application of the merger selection criterion  of
\citet{lot08} -- which had been established for the elliptical Petrosian
aperture -- is heavily biased by the strong aperture dependence of the 
Gini coefficient (see Sect.~\ref{sec:gini} and Fig.~\ref{fig:aperture}),
thus providing a straightforward explanation for the apparently very
different sensitivity of the CAS merger selection and the Gini/$M_{20}$
merger selection \citep[cf.][]{con08}.

The Gini coefficient was used by \citet{nei08} in their analysis of
intermediate-mass galaxies at redshift $z\approx 0.6$.
These authors find
good agreement of visual morphological classification with the dynamical
states of the galaxies, but point out that this correlation is not as
clear when using automated classification methods instead.
 With the Petrosian
aperture and a limiting magnitude of $i_{775}=23.5$ mag for GOODS images,
they are only weakly affected by $S/N$ effects: about 20\% of galaxies
with magnitudes around $i_{775}=23.5$ mag fall below the $S/N$
criterion.
The $S/N$ effect on the Gini coefficient can cause low-$S/N$
objects to be assigned to later galaxy types than they actually belong to (cf.\ Fig.~\ref{fig:class}),
or mergers to fall out of the merger regime as defined by \citet{lot08}
and used by \citet{nei08}. This could partly explain that the visually
assigned classes of \citeauthor{nei08} intermingle when using the
Gini/$M_{20}$ criterion, but only few of their
52 galaxies should be affected by this.
%

\citet{urr08} applied the Gini coefficient in their analysis of quasar
host galaxies, following the Gini calculation of \citet{lot04}. Their
$I_{814}$ HST/ACS imaging exposure times are 
$\lesssim 30$ times lower than the total exposure time of GOODS $i_{775}$,
corresponding to a difference in depth of $\lesssim 1.8$ mag. Since all
galaxies of their sample have magnitudes $I_{814}<21$ mag, their study
does not suffer from $S/N$ effects on the Gini values.

Based on COSMOS \citep{cosmos} HST/ACS data, \citet{cap07} and
\citet{cas08} applied the Gini coefficient to study the evolution of the
morphology-density relation, and to morphologically select faint AGN,
respectively. Both rely on the Gini calculation outlined in \citet{abr07},
using a ``quasi-Petrosian'' aperture. A magnitude limit of $I_{814}<24$
mag is used by \citet{cap07}, whereas \citet{cas08} adopt $I_{814}<24.5$
mag. Due to the larger pixel scale of the COSMOS data as compared to
GOODS, a galaxy would actually have
a slightly larger $S/N$ per pixel in COSMOS than in GOODS, despite the
shallower depth of COSMOS \citep{cosmoshst}.\footnote{Any
  possible effects of the lower spatial resolution of the COSMOS data on
  the Gini coefficient are not  considered here.}
%

Taking this into account, $\sim$30\% of galaxies with magnitudes close to
the limit of \citet{cap07} fall below the $S/N$ criterion established
above, and $\sim$15\% of all galaxies brighter than the limit would do
so.
These authors separated early and late-type galaxies according
to their Gini value, and therefore probably missed a certain fraction of
(faint) early types. Since mostly galaxies at higher
redshift should be affected, due to their lower overall $S/N$, the 
growth rate of the early-type fraction with cosmic time determined by
\citeauthor{cap07} may have been slightly overestimated.

For the study of \citet{cas08}, $\sim$35\% of galaxies close to
the magnitude limit do not meet the $S/N$ criterion, and $\sim$20\% of all
galaxies brighter than the limit do so. They applied the Gini coefficient to
select compact sources that are likely to be AGNs by their high Gini
value. Therefore, the authors might have missed a small fraction of galaxies at low
$S/N$ levels with Gini values close to the selection limit, which would
have entered their AGN candidate sample at larger $S/N$.

The fractions of $S/N$-affected galaxies given above for the study of
\citet{cap07} also apply to the COSMOS morphological classification
presented by \citet{sca07}. This leads to a certain number of (faint)
galaxies being classified as later types than they should, and is in
agreement with \citeauthor{sca07}'s estimate that $\lesssim30\%$ of faint
early types could be misclassified as disks or irregulars.

In their study of
the morphology of distant star-forming galaxies in COSMOS, \citet{zam07}
adopt a magnitude limit of $I_{814}<23$ mag, thus avoiding any significant
$S/N$ effect. Likewise, the selection of $I_{814}<22$ mag for COSMOS and
$i_{775}<22$ mag for GOODS ensured that no $S/N$ effect on the Gini values
affected the morphological analysis of \citet{celeste}, who presented a
preliminary method for identifying bars based on radial
Gini profiles.

\citet{abr07} stated that ``the Gini coefficient remains a surprisingly robust
statistic, even in the face of morphological $K$-corrections''. This was
based on their assessment of the robustness of the Gini coefficient
(computed within the ``quasi-Petrosian'' aperture), using  
an SDSS reference sample of nearby galaxies for their HST/ACS study of
distant early-type galaxies. The galaxies in both their HST/ACS sample and
their SDSS reference sample have a total $S/N>100$. When compared to the
GOODS data of the present study, indeed all galaxies with $S/N>100$ would
meet the $S/N$-criterion established above, in accordance with
\citeauthor{abr07}'s positive
conclusion on the performance of the Gini coefficient as morphology
estimator.

\section{Concluding remarks}

So far, most authors using the
Gini coefficient to describe galaxy morphology applied a reasonable
magnitude and/or $S/N$ selection that prevented their investigations from
being seriously affected by $S/N$ effects. However, I have
shown that the Gini coefficient depends strongly on the aperture within
which it is computed, and that it exhibits a strong dependence on $S/N$ below a
certain $S/N$ level (which, again, depends on the aperture).
 Therefore,
quantitative selection criteria involving the Gini coefficient -- such as
the merger selection of \citet{lot08} -- cannot be straightforwardly
applied to a different dataset than the one for which they have been
established \citep[cf.][]{con08}. Care needs to be taken with the
selection of aperture and limiting magnitude, as well as with the
comparison of calculated Gini values to those of other studies. From the
analysis presented here, the use of the Petrosian aperture is recommended.
Larger apertures are strongly disfavored due to their inclusion of many
faint pixels.

Typically, the total $S/N$ or the $S/N$ per pixel are quoted as
quantitative, but crude estimates of the ``quality'' of a galaxy's
image. From these average values, it can of course not be seen whether a given
galaxy image
actually consists of pixels with very different individual $S/N$ values, or
whether the $S/N$ of most pixels is similar --- which might make a huge
difference depending 
on the desired analysis. This pixel flux distribution is exactly what the
Gini coefficient, by definition, measures. One could thus conclude
somewhat ironically that the Gini coefficient could serve as a
sophisticated measure of the $S/N$ distribution within a galaxy's image, if it
did not depend so much on the galaxy's morphology.

\acknowledgements
    It is a pleasure to thank Ignacio Ferreras for stimulating discussions
    and initial help with the programming, and Vy Tran for useful and
    supportive comments. Jennifer Lotz is thanked  
    for clarifying some issues concerning the Gini calculation.
    I am supported within the framework of the Excellence Initiative
    by the German Research Foundation (DFG) through the Heidelberg
    Graduate School of Fundamental Physics (grant number GSC 129/1).
    Some preliminary investigations for this project benefitted from the
    facilities of 
    the former Astronomical Institute of the University of Basel.
    This research has made use of NASA's Astrophysics Data
    System Bibliographic Services.

 



\end{document}